\documentclass[12pt,usenames,dvipsnames,a4paper]{article}

\usepackage{amsmath,amsfonts,amssymb}
\usepackage{appendix}
\usepackage{caption}
\usepackage{cite}
\usepackage{color}
\usepackage{datetime}
\usepackage{float}
\usepackage{graphicx}
\usepackage[colorlinks,citecolor=blue]{hyperref}
\usepackage{indentfirst}
\usepackage[numbers,square,comma,sort&compress,merge]{natbib} 
\usepackage{subfig}

\def\a{\alpha}

\def\c{\chi}

\def\d{\delta}
\def\D{\Delta}

\def\eps{\varepsilon}
\def\f{\frac}

\def\G{\Gamma}

\def\l{\left}

\def\la{\langle}
\def\ra{\rangle}

\def\mc{\mathcal}

\def\m{\mu}

\def\r{\right}

\def\be{\begin{equation}}
\def\ee{\end{equation}}

\def\bea{\begin{eqnarray}}
\def\eea{\end{eqnarray}}

\def\ba{\begin{array}}
\def\ea{\end{array}}

\def\bc{\begin{center}}
\def\ec{\end{center}}

\def\bl{\begin{flushleft}}
\def\el{\end{flushleft}}

\def\br{\begin{flushright}}
\def\er{\end{flushright}}

\def\bi{\begin{itemize}}
\def\ei{\end{itemize}}

\def\bt{\begin{tabular}}
\def\et{\end{tabular}}

\begin{document}

\thispagestyle{empty}

\bc
\Large {\bf Inconsistencies of higgsplosion}
\ec

\bc
{\bf A.~Monin}
\ec

\bc
{\small Department of Theoretical Physics, University of Geneva} \\
{\small 24 quai Ernest-Ansermet, 1211 Geneva} \\
\ec

\bc

\texttt{\small alexander.monin@unige.ch} 

\ec

\abstract{

It is shown that higgsplosion scenario is impossible within local QFT framework.

\section{Introduction}

The standard, if not the only general analytic, tool for dealing with weakly coupled systems is perturbation theory. Producing an asymptotic expansion\footnote{This is a generic situation.} perturbation theory is reliable only up to a certain order, which is defined by the coupling. In cases when this order is high enough that does not present a problem, for having several subleading contributions anyway exhausts experimental accuracy. However, it is in general not clear how to deal with cases when perturbation theory breaks down very early. This is precisely the situation when processes involving a large number of particles (bosonic states) are considered.

Interest to such processes was initially sparked by computations in electro-weak theory involving instantons (see~\cite{Voloshin:1994yp,Rubakov:1995hq,Libanov:1997nt} for a detailed review). Later a more tractable model, namely the one of a scalar field with $\lambda \phi^4(x)$ self interaction (with or without spontaneous symmetry breaking) was considered, leading to several closed form results. The main object of interest in this model is the matrix element of the field operator between the vacuum and an $N$-particle state 
\be
\la N | \phi (x) | 0 \ra.
\label{eq:formFactor}
\ee

Summing all tree diagrams the leading order result for the amplitude at threshold (all final state particles have zero spatial momentum and energy equal to their mass $m$) was obtained in~\cite{Voloshin:1992mz,*Argyres:1992un}. A more convenient technique, employing classical equations of motion, developed in~\cite{Brown:1992ay} allowed to reproduce the leading order results and was later used to compute one-loop corrections~\cite{Voloshin:1992nu,*Smith:1992rq}. Schematically the behavior of (\ref{eq:formFactor}) was found to be
\be
\la N | \phi (x) | 0 \ra_\text{thresh}^\text{1-loop} \sim N! \lambda ^ {N/2} (1+ c \lambda N^2),
\label{eq:ffLeading}
\ee
with $c$ a constant. This result clearly indicates the breaking of perturbation theory for $N \sim 1/ \sqrt{\lambda}$. Applying a non-perturbative method~\cite{Gorsky:1993ix} for computing the matrix element at threshold resulted in\footnote{The result is valid insofar as $\lambda N \gg 1$.}
\be
\la N | \phi (x) | 0 \ra_\text{thresh}^\text{exact} \sim N! \,  \exp \l ( \tilde c \, N^{3/2} \sqrt{\lambda} \r ), ~~ \tilde c \geq 0.
\label{eq:nonPertAmp}
\ee
What is remarkable is the persistence of the factorial growth of the amplitude with the number of particles $N$. That could potentially lead to the unconstrained growth of the corresponding probability and the violation of unitarity. However, it is generally expected~\cite{Gorsky:1993ix,Rubakov:1995hq,Libanov:1997nt}, although remains an open question (see discussion below), that above threshold the amplitude develops a rapidly decaying form factor so that unitarity constraints~\cite{Froissart:1961ux,*Martin:1965jj,*Yndurain:1972ix} are satisfied.

Another quantity, related to the amplitude (\ref{eq:formFactor}) is the $N$-particle contribution to the spectral density
\be
\rho(E,N) = \sum _f | \la f | P(N,E) \phi (x) | 0 \ra | ^2
\label{eq:spectralNGen}
\ee
with $P(N,E)$ being a projector on $N$-particle states with energy $E$. A general method, based on singular classical solutions, for computing\footnote{In fact the method was formulated for an arbitrary operator $\mc O(x)$.} (\ref{eq:spectralNGen}) was formulated in~\cite{Son:1995wz}. The validity of the method for $\lambda N < 1$ was checked in several papers~\cite{Bezrukov:1995ta,*Bezrukov:1997xg,*Bezrukov:1999kb}, in particular results of explicit perturbative computations were reproduced.

However, only recently computations for the case of $\lambda N \gg 1$ and $E> N m$ were performed~\cite{Khoze:2017uga,*Khoze:2017lft,Khoze:2017ifq,Khoze:2017tjt}. It is claimed that for sufficiently small average kinetic energy per particle
\be
\eps = \f {E-m} {N m} \ll 1,
\ee
the result is given by
\be
\rho (E,N) \sim \eps^ {\f {3} {2} N} \exp \l ( \tilde c \, N^{3/2} \sqrt{\lambda} \r ) \l [ 1 + O (N \log N) \r ], ~~ \text{with} ~~ 
\tilde c = \mathrm{const}.
\label{eq:spectralN}
\ee
It is clear that for any fixed $\eps \neq 0$ and sufficiently large $N \gg 1$ the exponential term is dominating. In other words $N$-particle contribution to the spectral density close to threshold is exponentially large for large $N$ and fixed $\eps$
\be
\rho \big ( Nm (1+ \eps), N  \big ) \underset{\substack{N \to \infty \\ \eps = \text{fixed}}} {\sim} \exp \l ( \tilde c \, N^{3/2} \sqrt{\lambda} \r )\underset{\substack{N \to \infty \\ \eps = \text{fixed}}} {\to} \infty.
\label{eq:spectralNDiv}
\ee
Using this fact the authors of~\cite{Khoze:2017uga} conclude that the Feynman propagator decays exponentially for large energies, leading to a completely different, from what is usually expected, UV behavior of the theory.

Below we discuss in detail why the scenario with exponentially growing spectral density (\ref{eq:spectralN})
is impossible within the framework of local QFT. In short, locality puts constraints on the behavior of correlators at infinity in momentum space, which are not satisfied by (\ref{eq:spectralN}). It means that contrary to what is stated in~\cite{Khoze:2017uga, Khoze:2017lft}, $\lambda \phi^4$ model is not UV complete. Although this is clearly not new, what seems rather unusual is that the necessary cutoff is not $\sim m \exp \l ( 16\pi^2 / \lambda \r )$, but rather can be unexpectedly low
\be
\Lambda \gtrsim | m \log^2 \eps_\text{max} |,
\label{eq:effCutOff}
\ee
where $\eps_\text{max} <1$ is the maximal value of kinetic energy per particle, for which the computation (\ref{eq:spectralN}) can be trusted.

Moreover, with the claim that the Feynman propagator decays exponentially at infinity in Minkowski region, analyticity and unitarity imply its exponential growth in Euclidean region, which renders arguments of~\cite{Khoze:2017lft, Khoze:2017uga} about the appearance of a rather low effective cutoff in loop integrals invalid. We show below that this claim is actually in contradiction with (\ref{eq:spectralN})  rather than its consequence. We would like to point out that the value of the would be cutoff (around $200 m$) quoted in~\cite{Khoze:2017lft, Khoze:2017uga} depends not only on the coupling constant $\lambda$ but also on the domain of validity of (\ref{eq:spectralN}), as is evidenced from (\ref{eq:effCutOff}).

\section{The Weinberg theorem}

In this section we show that the standard assumptions (unitarity, locality and analyticity) about QFT set a bound on the behavior of the Feynman propagator at infinity in momentum space, known as the Weinberg theorem. Even though it is a textbook material (see~\cite{Weinberg:1995mt}), we keep it for the sake of being self consistent. To derive the K\'all\'en-Lehmann representation we start from writing down the spectral decomposition for the two-point Wightman function of a scalar operator
\be
W_\mc O (x) = \la \mc O (x) \mc O (0) \ra.
\label{eq:Wightman}
\ee
We choose a basis in the Hilbert space formed by eigenstates of momentum
\be
P^\m | n \ra = p_n^\m | n \ra
\ee
with $p_n = (\sqrt{m_n^2+\vec p ^ 2}, \vec p)$ and $m_n^2$ being the eigenvalue of the Casimir operator 
\be
P_\m P^\m | n \ra = m_n^2 | n \ra.
\ee
Using the completeness relation
\be
\sum _n | n \ra \la n | = 1,
\ee
and the way operators $P_\m$ act on fields
\be
\mc O (x) = e ^ {i P x} \mc O (0) e^ {-i P x},
\ee
we obtain
\be
W_\mc O(x) = \sum _n | \la 0 | \mc O (0) | n \ra | ^ 2 e^{-i p_n x},
\label{eq:Wightman_Completeness}
\ee
Writing explicitly the summation over the three-momentum $\vec p$ as an integral
we may rewrite (\ref{eq:Wightman_Completeness}) as
\be
W _\mc O(x) = \int \f{ds} {2\pi} \f {d^3p} {(2\pi)^3} \f {e^{-i p_s x}} {2 E_{s,\vec p}} \rho_\mc O(s),
\label{eq:Wightman_spectral}
\ee
with $p_s = (E_{s,\vec p},\vec p)$, $E_{s,\vec p} = \sqrt{s+\vec p ^ 2}$, and the spectral density of the operator $\mc O(x)$ given by
\be
\rho_\mc O(s) = 2 \pi \sum_{\substack{n: \\ \vec p =\text{fixed}}} | \la 0 | \mc O(0) | n \ra | ^ 2 \d (s-m_n^2).
\label{eq:spectral}
\ee
It is clear from (\ref{eq:Wightman_spectral}) that the Wightman function (\ref{eq:Wightman}) exists only in the sense of a distribution. Usually in local QFT it is assumed to be a tempered distribution, i.e. defined on the space of rapidly decaying (Schwartz) test functions. That imposes constraints on the behavior of the propagator and spectral density at infinity, namely, it cannot grow faster than a polynomial at $s \to \infty$. Indeed, for a test function from the Schwartz space $g (x) \in \mathcal{S}(\mathbb{R}^4)$, we have
\be
\int W_\mc O(x) g(x) d^4 x = \int \f{ds} {2\pi} \f {d^3p} {(2\pi)^3} \f {\tilde g (p_s)} {2 E_{s,\vec p}} \rho_\mc O(s),
\ee
with $\tilde g(p)$ being the Fourier transform of $g(x)$. The condition for the integral to exist is that the spectral density be bounded by a polynomial. At the same time there is no bounds on how fast the Fourier transform 
$\tilde W_\mc O (p)$ can decay at infinity\footnote{Although if it decays sufficiently fast, the corresponding field is necessarily the generalized free one~\cite{Vasilev:1965, *Baumann:1984cq}.}.

The situation is different for the time ordered two-point function
\be
D_\mc O(x) = \la T \mc O(x) \mc O(0) \ra,
\ee
whose Fourier transform $\tilde D_\mc O(p^2)$ cannot decay faster than $p^{-2}$. Indeed, the analogue of (\ref{eq:Wightman_spectral}) has the following form
\be
\tilde D_\mc O (p^2) = \int \f{ds} {2\pi}\f {i} {p^2-s+i \eps} \rho_\mc O(s),
\label{eq:time_Ordered_spectral}
\ee
which should be understood as having a number of subtractions if $\tilde D_\mc O(s) \underset{s\to \infty}{\not\to}0$. It is important, though, that the number of subtractions, due to the assumption of temperedness, is always finite. After $K$ subtractions one has
\be
\tilde D_\mc O (p^2) = P_{K-1} (p^2) + p^{2K} \int \f{ds} {2\pi}\f {i} {p^2-s+i \eps} \f{\rho_\mc O(s)}{s^K},
\label{eq:spectral_subtraction}
\ee
with $P_{K-1}(p^2)$ being a polynomial of degree $K-1$ with real coefficients. Expressions (\ref{eq:time_Ordered_spectral}) or (\ref{eq:spectral_subtraction}) allow to analytically continue the $\tilde D_\mc O (p^2)$ in the whole complex plane with a cut, starting at the two-particle threshold. The discontinuity across the cut is obviously given by
\be
2\mathrm{Im} (i \tilde D_\mc O(p^2)) = \rho_\mc O(p^2).
\label{eq:Unitarity}
\ee
And a direct consequence of this analytic continuation is that the fastest the propagator $\tilde D_\mc O(p^2)$ can decay at infinity in any direction is $p^{-2}$.

\section{Nonlocality and higgsplosion scenario}

Correlators being tempered distributions implies locality of the underlying QFT. However, that does not exhaust all the realizations of  local QFT. In fact generalizations of the condition that the propagator (spectral density) be bounded by a polynomial were studied in a number of papers~\cite{Meiman:1966,Guttinger:1966,Jaffe:1967nb,Efimov:1967pjn,Voronov:1967zz,*Iofa:1969fj,*Iofa:1969ex}. For exponentially growing spectral density (\ref{eq:Wightman_spectral}) or (\ref{eq:time_Ordered_spectral}) do not exist as tempered distributions. In this case for them to make sense a different space of test functions should be considered. These test functions should decay faster then the exponent of a certain power $\exp \l ( {-A |p^2|^{\a}} \r )$. 

It was shown in~\cite{Meiman:1966,Guttinger:1966,Jaffe:1967nb} that only for $\a <1/2$ the system can be considered local. In short, fast decaying in momentum space test functions correspond to analytic (entire) functions in coordinate representation. The latter ones cannot have a finite support, therefore, they cannot be used to localize the system. Let us consider the following example. Exponentially growing spectral density necessitates an infinite number of subtractions in (\ref{eq:spectral_subtraction}), leading to
\be
F(x)=\sum_{n=0}^\infty a_n \Box^{n} \d(x),
\label{eq:genFunc}
\ee
in coordinate representation instead of a polynomial in (\ref{eq:time_Ordered_spectral}). The question is, when does (\ref{eq:genFunc}) correspond to a local functional? For the purpose of illustrating the idea we consider a one-dimensional situation. In this case for a test function $g(x)$ the functional becomes 
\be
F[g]= \sum_{n=0}^\infty a_n g^{(2n)}(0),
\ee
which can be rewritten, using the Cauchy integral to express $g^{(2n)}(0)$, as
\be
F[g]= \sum_{n=0}^\infty a_n \f{(2n)!} {2 \pi i} \oint \f{g(z)} {z^{2n+1}} dz = 
\f{1} {2 \pi i} \oint \f{g(z)} {z} g_F (z) dz,
\label{eq:Cauchy}
\ee
with the following definition
\be
g_F (z) = \sum_{n=0}^\infty a_n \f{(2n)!}{z^{2n}}.
\label{eq:charSeries}
\ee
The function $g_F(z)$ is well-defined provided the radius of convergence
\be
R= \lim_{n\to \infty} \l ( (2n)! \, |a_n| \r ) ^ {-1/n},
\ee
of the series (\ref{eq:charSeries}) is not zero. In this case the contour of integration in (\ref{eq:Cauchy}) cannot be made smaller then $R^{-1}$, which serves as a measure of non-locality. If it happens that the radius of convergence is infinite, then the contour of integration can obviously be shrunk to zero and one can say that in this case the functional is local (with the support at $x=0$). For the latter to be realized it is needed that
\be
\l ( (2n)! \, a_n \r ) ^ {1/n} \underset{n \to \infty } {\to} 0,
\ee
or in other words the Fourier transform
\be
\tilde F(p^2)=\sum_{n=0}^\infty a_n p^{2n}
\ee
should be an entire function of order (see Appendix) less then $1/2$, i.e. 
\be
\tilde F (p^2) \sim e^{\tilde A |p^2|^\a}, ~~ \text{with} ~~ 0 < \a < \f {1} {2},~~ \text{and} ~~ \tilde A = \mathrm{const}>0.
\label{eq:localityCondition}
\ee

A more intuitive picture results form considering a theory of the generalized free field $\Phi(x)$ of dimension $\D_\Phi$, which (at least intuitively) we would like to call local. The propagator of a monomial $\Phi^k(x)$ with finite power $k$ needs only a finite number of subtractions, since
\be
D_{\Phi^k} (x) \underset{x^2 \to 0}{\to} \f {k! A^k} {(x^2)^{k \D _\Phi}}.
\ee
Hence, those can be defined as tempered distributions. At the same time for an exponential 
\be
\la T e^{{\Phi(x)} / {f}} e^{{\Phi(0)} / {f}} \ra = e^{D_\Phi (x)/f^{2\D_\phi}} \underset{x^2 \to 0}{\to} 
\exp \l [ \f {A} {(x^2 f^2)^{\D _\Phi}} \r ], ~~ f = \mathrm{const},
\ee
even though infinite number of subtractions is needed, the growth of the Fourier transform at infinity is bounded by 
$\exp A (p^2)^\a$, with $\a < 1/2$, which can be checked easily using the saddle point approximation. Generally, considering an operator given by a formal series defined by an entire function
\be
\mc O (x) = \sum _{k=1} ^\infty C_k \Phi ^ k (x),
\ee
we find
\be
D_\mc O (x) = \sum _{k=1} ^\infty  C_k^2 \, D_{\Phi^k}(x) \underset{x^2 \to 0}{\to}
\sum _{k=1} ^\infty  k! C_k^2 \f {A^k} {(x^2)^{k \D _\Phi}}.
\label{eq:genCorrEnt}
\ee
For the series above to make sense its radius of convergence should be infinite, meaning that 
$(k! C^2_k)^{1/k}\underset{k \to \infty} {\to} 0$, which in turn implies (see Appendix) that the analogue of (\ref{eq:localityCondition}) for the Fourier transform of (\ref{eq:genCorrEnt}) is satisfied.

Now, according to~\cite{Khoze:2017uga, Khoze:2017lft, Khoze:2017ifq, Khoze:2017tjt} the spectral density grows exponentially with energy. Indeed,
\be
\rho(p^2) \geq \sum _N \rho(\sqrt{p^2},N) \geq \rho (\sqrt{p^2},N_\text{max}),
\label{eq:spectralDensity}
\ee
with $N_\text{max}$ the biggest kinematically allowed number of particles. Hence, it is evident from (\ref{eq:spectralNDiv}) that at least for a sequence of points $\sqrt{p^2} = N m (1+\eps)$ with $\eps \ll 1$ the spectral density diverges exponentially. That, among other things, implies that the Feynman propagator also diverges exponentially in Minkowski region as it follows from (\ref{eq:Unitarity}), whose derivation does not have to rely on (\ref{eq:time_Ordered_spectral}), for it is the consequence of unitarity\footnote{Considering a coupling $g \chi ^2 \phi$ and using the optical theorem for $\c \c\to \c \c$ leads immediately to (\ref{eq:Unitarity}).}. As it follows from (\ref{eq:spectralNDiv}) and (\ref{eq:spectralDensity}) the corresponding order of the Feynman propagator is $\a \geq 3/4$, thus, eliminating the possibility to realize the scenario within local QFT.

\begin{figure}[h]
\bc
\includegraphics[width=5cm]{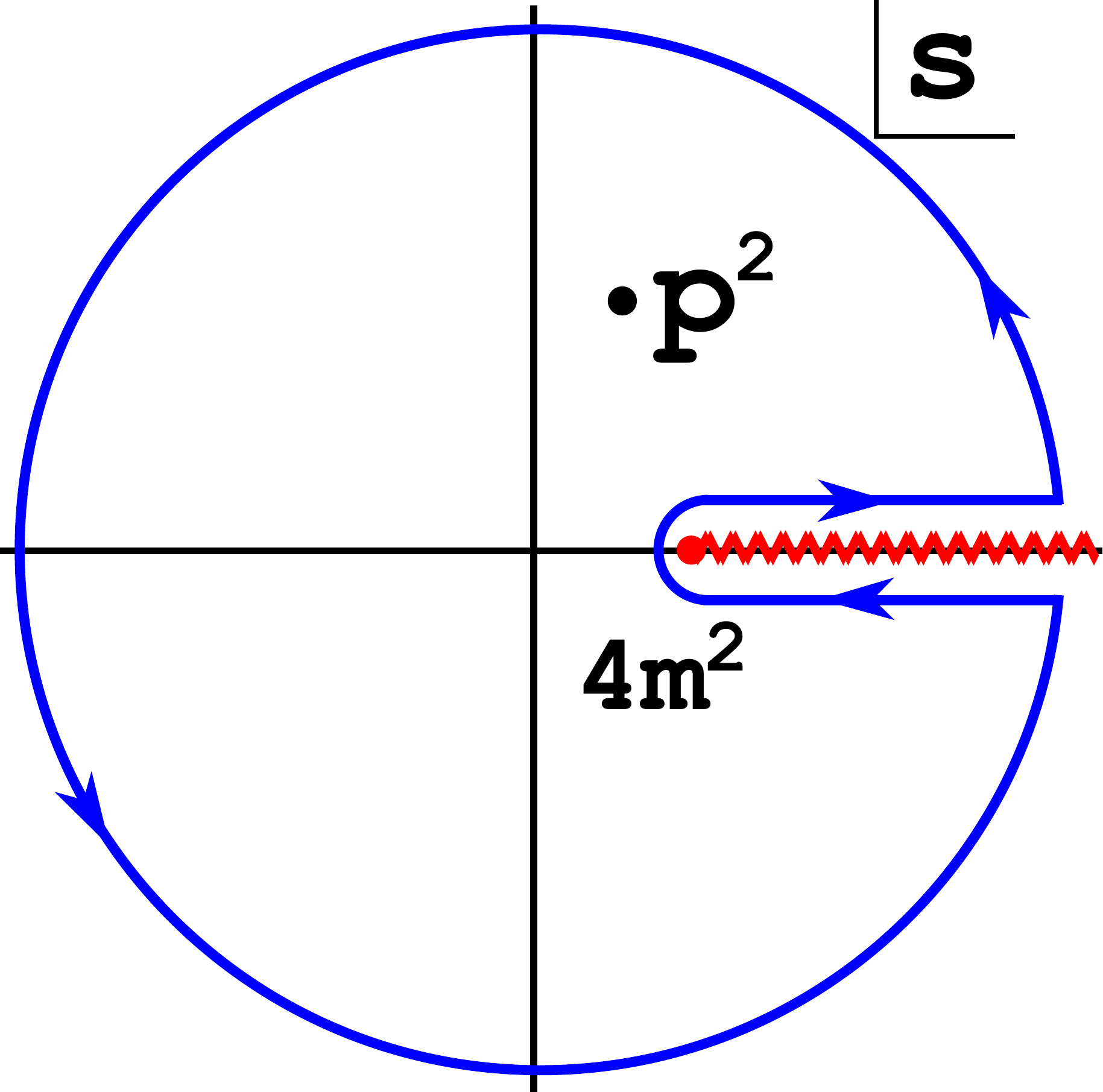}
\ec
\caption{\label{fig:contour} Integration contour.}
\end{figure}

\section{Comments about decaying propagator} 

It should be noted that in~\cite{Khoze:2017uga, Khoze:2017lft} it is claimed that the propagator in Minkowski region, despite  (\ref{eq:spectralDensity}), does not grow but rather decays exponentially
\be
\tilde D (p^2) \underset{p^2\to + \infty} {\sim} O(e^{-C (p^2)^{3/4}}), ~~ C>0.
\label{eq:decayFeynman}
\ee
This claim obviously contradicts the Weinberg theorem. In addition, as we argued above, unitarity alone guarantees (\ref{eq:Unitarity}), therefore, the spectral density should decay provided  (\ref{eq:decayFeynman}) is true, which would in turn contradict (\ref{eq:spectralDensity}).

Moreover, analytic continuation of the propagator to Euclidean region\footnote{Note that the analytic properties of the propagator are tacitly assumed in~\cite{Khoze:2017tjt} to be the standard ones: the only singularity (modulo poles) is the cut on the real axis starting at threshold, Fig.~\ref{fig:contour}.} is made in~\cite{Khoze:2017uga} somewhat hastily. It is simply stated that for $p^2\to - \infty$ the propagator also decays exponentially. That cannot be the case, since analytic continuation of (\ref{eq:decayFeynman}) along a path in the upper half plane implies that the propagator at $p^2 \to -\infty$ should grow at least as $\exp \l ( \f{C} {2} |p^2|^{3/4} \r )$.

\section{Conclusion}

Axiomatic approach to QFT, despite being cumbersome at times, provides us with robust and model independent predictions. Even though it is hard (if not impossible) to guarantee that all the assumptions going into theorems of axiomatic QFT are met in specific cases, for we mostly use perturbative analysis, the consequences of those theorems may be used at least as a guide when studying concrete realizations of QFT. Deviation from these predictions signals that something unusual is happening or that the result should be taken with a grain of salt.

There are several claims in~\cite{Khoze:2017uga, Khoze:2017lft} about $\lambda \phi^4$ theory that certainly raise a flag. Exponential decay of the Feynman propagator at infinity in momentum space definitely contradicts the Weinberg theorem. Moreover, it is inconsistent with the suggested UV behavior of the spectral density, provided unitarity and analyticity are intact: there is no energy scale for which the propagator decays exponentially in Minkowski region, on contrary it grows together with exponentially growing number of new states.

At the same time exponential growth of the spectral density also evokes scepticism. As we demonstrated such a behavior indicates that the underlying theory is effective with (possibly) very low cutoff, which is not what is usually expected, based on perturbative analysis. We conjecture that in fact the formula for the $N$-particle contribution to the spectral density is only valid (if at all) for extremely small kinetic energy per particle, thus, pushing the would be cutoff higher. However, only alternative derivations of the asymptotic behavior of the spectral density, for instance from the lattice, could unambiguously resolve the issue.

\section{Acknowledgements}

I am thankful to R.~Rattazzi, M.~Shaposhnikov, F.~Bezrukov and S.~Sibiryakov for useful discussions. The work is supported by ERC-AdG-2015 grant 694896.

\appendices

\section{Entire functions}

An entire function can be defined by its Taylor series
\be
g(z) = \sum_{n=0}^\infty a_n z^n,
\ee
with infinity radius of convergence, meaning that 
\be
a^{1/n}_n \underset{n\to \infty} {\to} 0.
\ee
An entire functions is of a finite order if for $|z| \to \infty$ there exists $A>0$ for which
\be
g(z) = O (e^{|z|^A}).
\ee
The lower bound $\a$ of numbers $A$ is called the order of this function. The order $\a$ can be computed from
\be
\a = \lim_{n\to \infty} \sup \f {n \log n} {-\log |a_n|}.
\ee
As an example, it is easy to show by approximating the function with the largest term in the series, that the following functions are of order $\a$
\be
\sum_{n=0}^\infty \f{z^n} {\G(n/\a+1)}, ~~\text{or} ~~ \sum_{n=0}^\infty \f{z^n} {[\G(n+1)]^{1/\a}},
\ee
with $\G(z)$ being the Euler gamma function.

It is proven in~\cite{Efimov:1967pjn,Efimov:1977} that the Fourier transform (understood as a distribution) of an entire function $D(x^2)$ of order $\a$ (at zero)
\be
D(x^2) \underset{x^2 \to 0} {=} O \l ( \exp {\f{C}{(x^2)^\a}} \r ),
\ee
is itself an entire function $\tilde D (p^2)$ of order $\tilde \a = \a / (2 \a +1)$
\be
\tilde D (p^2) \underset{p^2 \to \infty} {=} O \l ( \exp \tilde C(p^2)^{\tilde \a} \r ).
\ee
Formally, one can see that by using a saddle point approximation to compute $\tilde D (p^2)$ for large $p^2$.

\section{Analogy}

In this section instead of rederiving the computation~\cite{Khoze:2017uga,*Khoze:2017lft,Khoze:2017ifq,Khoze:2017tjt}
we illustrate the idea of the non-perturbative computation by considering a prototypical example exhibiting the difficulties of perturbative analysis. Suppose we want to compute the following one-dimensional integral
\be
F(g,N) = \int_0^\infty x^{2N} \exp \l ({-\f{x^2}{2}-\f{g x^4}{4}} \r ), ~~ N \in \mathbb Z.
\label{eq:1dIntegral}
\ee
Usual perturbation theory corresponds to formally expanding the integrand in powers of $g$. Doing that we find (compare with (\ref{eq:ffLeading}))
\be
F(g,N) = \sum_{k=0}^\infty F_k g^k = 2^{N-\f{1}{2}}\G \l (N+\f{1}{2} \r ) \l [ 1 -\f{g}{4} (2N+3)(2N+1) +\dots \r ].
\label{eq:prototypeSeries}
\ee
Coefficients $F_k$ grow factorially with $k$, namely
\be
F_k \sim \f{(N+2k)!} {k!},
\ee
therefore, the radius of convergence of (\ref{eq:prototypeSeries}) is zero, it is an asymptotic series, accurately representing the function $F(g,N)$ for $gN \ll 1$. One way to find $F(g,N)$ is to use Borel summation, producing
\be
F(g,N) = \f{1} {g} \int_0^\infty dt e^{-t/g} \mc B F(t,N) =
g^{-\l ( \f {N} {2} + \f {1} {4}\r )} \G \l (N+\f{1}{2} \r ) U \l (\f{N}{2}+\f{1}{4},\f{1}{2},\f{1}{4g} \r ),
\label{eq:integrationResult}
\ee
with Borel image defined as
\be
\mc B F(t,N) = \sum_{k=0}^\infty \f{F_k} {k!} t^k,
\ee
and $U(a,b,z)$ being confluent hypergeometric function. One can check that computing the integral (\ref{eq:1dIntegral}) exactly indeed results in (\ref{eq:integrationResult}).

Even without knowing the exact result, it is evident from (\ref{eq:prototypeSeries}) that perturbation theory breaks down for 
$g N^2 \gg 1$. The way to obtain the behavior of the integral in a non-perturbative regime, i.e. for $N \gg 1$, is to use a different saddle point.  For the case at hand, instead of expanding around $x=0$, which is the stationary point of the exponential in (\ref{eq:1dIntegral}), we look for a new saddle taking into account also $x^{2N}=e^{2N \log x}$ term. As a result we obtain
\be
F(g) \underset{gN \gg 1}{=}\f{\sqrt{\pi}} {(8 g N)^{1/4}}\exp \l [ \f {N} {2} \l ( \log \f {2 N} {g} - 1 - \sqrt{\f{2} {g N}}\r ) \r ] \l [ 1 + O \l ( \f{1}{\sqrt{g N}} \r ) \r ],
\ee
which is another asymptotic representation of (\ref{eq:integrationResult}), valid in a different regime, $g N \gg 1$. In the context of computing the amplitude (\ref{eq:formFactor}) at threshold the new saddle was found in~\cite{Gorsky:1993ix}, leading to (\ref{eq:nonPertAmp}). While in~\cite{Khoze:2017uga} the same saddle plus corrections due to nonzero kinetic energy resulted in (\ref{eq:spectralN}).

\bibliographystyle{utphys}
\bibliography{higgsplosion}{}

\end{document}